\UseRawInputEncoding
\documentclass[prl, twocolumn]{revtex4-2}
\usepackage{amssymb}
\usepackage{amsmath}
\usepackage{epsfig}
\usepackage{bm}

\begin{document}

\title{
Analytical model for non-linear magnetotransport  in viscous  electron fluid
}

\author{  P. S. Alekseev and M. A. Semina }

\affiliation{  Ioffe  Institute,  194021  St.~Petersburg, Russia }

\begin{abstract}

We develop an analytical theoretical model for non-linear hydrodynamic  magnetotransport of two-dimensional (2D) electron fluid with strong
pair correlations in the electron dynamics.  Within  classical kinetics of 2D electrons, such correlations are described as
subsequent ``extended'' collisions of the same electrons, temporarily joined in pairs. Corresponding correlation-induced
retarded terms in the fluid dynamic equations can be described for slow flows as the dependence of the electron fluid viscosity
on the flow velocity gradient, that is  Non-Newtonian behavior of the fluid.  We  analytically calculate flow profiles in long samples in a stationary
highly non-linear regime and the corresponding  magnetoresistance. Pair correlations lead to a characteristic non-monotonic dependence of
 the differential resistance on magnetic field. We compare our results with experimental data on non-linear magnetotransport of
high-purity GaAs quantum wells; we conclude that our model can be responsible for a part of the observed features of the differential
 magnetoresistance.

\end{abstract}

\maketitle

{\em 1. Introduction.} In solid-state systems with low densities of defects the inter-particle interaction can lead to formation
 of a viscous  fluid  of conduction electrons  and realization of the  hydrodynamic regime of electric transport.  About ten years ago,
this was experimentally demonstrated  for  2D layered metal PdCoO$_2$~\cite{Moll}, single-layer
 graphene~\cite{graphene_1,Levitov_et_al,Polini_Geim},  and GaAs quantum wells  Ref.~\cite{je_visc,Gusev_1,Keser,recentest_,f1}.
This regime was detected by a specific dependence of the resistance on the sample width~\cite{Moll},  by the observation of a negative
nonlocal resistance related to whirlpools~\cite{graphene_1,Levitov_et_al}, by the the decrease of the resistance with temperature
(the Ghurzhi effect)~\cite{Gusev_1}. Direct space-resolved  observation of  profiles of the Hall electric field and the hydrodynamic
velocity  in graphene stripes  was performed~\cite{graphene_4,graphene_5}.  The giant negative
magnetoresistance is a characteristic sign of the hydrodynamic electron transport in magnetic field. It was observed
in high-quality GaAs quantum wells~\cite{exps_neg_1,exp_GaAs_ac_1,exps_neg_2,exps_neg_3,exps_neg_4,recentest_,exp_GaAs_ac_1}
and  explained~\cite{je_visc} as a result  of  formation of a  2D electron viscous  fluid~\cite{Gurzhi_Shevchenko}.  Later,
a very similar huge negative magnetoresistance was detected in  other  high-quality conductors: 2D metal PdCoO$_2$~\cite{Moll},
the  3D Weyl semimetal WP$_2$~\cite{Gooth}, and single-layer  graphene~\cite{graphene_4,graphene_5}, for which  other  evidences of
 the electron  hydrodynamics were  observed~\cite{graphene_1,Polini_Geim,graphene_4,graphene_5}.

Other non-trivial  effects of electron hydrodynamics in moderate magnetic fields were studied in subsequent works (see, for example,
Refs.~\cite{graphene_3,vis_res_0,vis_res_1,vis_res_2,Semiconductors,a,Holder,c}).   In Ref.~\cite{graphene_3} the non-diagonal
(Hall) viscosity  coefficient of 2D electrons, appearing in magnetic field,  was measured in graphene  samples of complex shape.
It was shown that the diagonal and non-diagonal high-frequency viscosity coefficients of 2D electrons exhibit  a single resonance
at the twice  cyclotron frequency~\cite{vis_res_0,vis_res_1}. In the case of a strongly non-ideal fluid, such ``viscoelastic''
resonance manifests itself  via excitation  of the  sher stress waves~\cite{vis_res_2,Semiconductors}.
 In Refs.~\cite{a,Holder,c} were studied the transition from the ballistic to the hydrodynamic regime of 2D electron transport
 in high-quality samples with the increase of magnetic field and the coexistence of the hydrodynamic and the ballistic contributions
in the Hall effect in a Poiseuille flow.

Experimental studies of the  non-linear regime of 2D electron transport  in ultra-pure samples of GaAs quantum wells were performed~\cite{non-lin_hydr_1,non-lin_hydr_3,non-lin_hydr_4}. The evolution of the giant negative magnetoresistance with the increase
of the current was examined~\cite{non-lin_hydr_1}.   The formation of  the hydrodynamic regime with the   increase
of the current, inducing the growth of the interparticle scattering,   was observed in Refs.~\cite{non-lin_hydr_3,non-lin_hydr_4}.
 It was theoretically shown~\cite{n-N} that the long-lived modes in 2D degenerated electrons in zero magnetic field, related to
the odd harmonics of their distribution function, lead to a non-Newtonian behavior of the electron fluid: formation of
complex modes with   non-linear dependencies of  their frequencies on the viscosity.

In this work we construct a theoretical model for non-linear hydrodynamic magnetotransport in a 2D viscous electron fluid.  We account
the non-linear effects  in the retarded relaxation of the fluid  shear stress, those are induced by the pair correlations
 in cyclotron  rotation of electrons [see Figs.~(a,b)].  We apply the phenomenological model of Ref.~\cite{new} describing such pair
correlations within the macroscopic fluid dynamics equations. For slow  Poiseuille  flows, the non-linear retarded terms in these equations
can be described as the dependence of the viscosity on the velocity gradient. As a result, the electron fluid turns out to be a pseudoplastic
or dilatant non-Newtonian fluid, depending on magnetic field.  At some diapason of flow parameters, ``strongly non-Newtonian behavior''
appears:   the viscosity becomes even a non-monotonic function of the velocity gradient. We analytically calculate the velocity profiles
and the corresponding differential magnetoresistance, with  turns out to be strong and non-monotonic.  We  compare our predictions with
experimental results of Ref.~\cite{non-lin_hydr_1} on magnetotransport of high-quality  GaAs quantum wells,  where the hydrodynamic
 regime was apparently realized.  Some of   the properties of the observed  magnetoresistance (a non-monotonic dependence on magnetic
field) can be explained by our results. We mention that other properties can be explained by  the effect of heating of electrons.

\begin{figure}[t!]
\centerline{\includegraphics[width=.99 \linewidth]{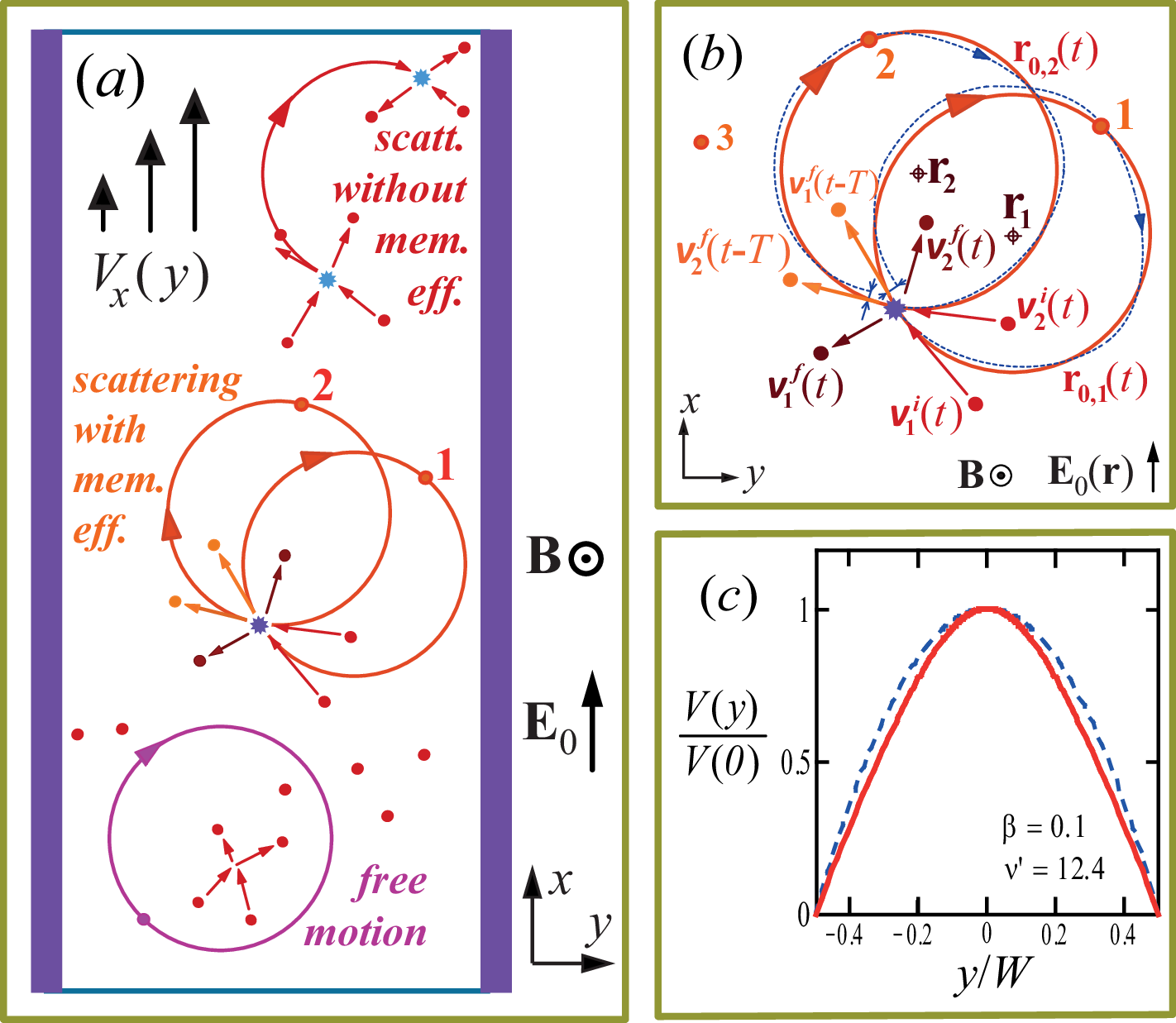}}
\caption{
($a$):
Long sample with 2D electrons in magnetic and electric  fields; classical-mechanics sketches  of different scattering events
leading to formation of the  viscous electron fluid flow with pair correlations.
($b$):
Extended collision of two electrons ``1'' and ``2''. The initial and the final velocities of the two electrons in two recollision events
are shown. The solid red lines are the circle unperturbed trajectories in magnetic field, while dashed blue lines are the trajectories
 perturbed by the forces from  the inhomogeneous electron  flow.
($c$):
The velocity profile (red curve) of the electron fluid flow calculated within the proposed mechanism of non-linearity due to
the pair correlation, the case of a strong non-linearity   ($ \nu'=12.4$) and  small classical magnetic field ($\beta=0.1$).
 The dashed blue line is the parabolic profile  with the same amplitude as the red curve, shown for demonstration of the non-parabolic
shape  of the red curve.
  }
\end{figure}

 {\em 2. Model for non-linear magnetotransport of 2D electron fluid. }
In Refs.~\cite{el,el2,el3} a hydrodynamic-like viscoelastic  model of ac flows of a 2D electron fluid, based on the Fermi-liquid
theory, was developed. Within a similar approach, in Refs.~\cite{vis_res_2,Semiconductors}  a  model of flows of a highly-correlated viscous
 2D electron fluid  in classical magnetic fields was constructed.
In many works (see, for example, Refs.~\cite{S1,S12,S2,S3,S4,S5,S6,S7,Beltukov_Dyakonov})
memory effects in magnetotransport of non-interacting 2D  electrons,
 scattering on defects in a sample, were theoretically studied.
In Ref.~\cite{new}, on base of the approaches of Refs.~\cite{Beltukov_Dyakonov} and~\cite{el,Semiconductors}
 a phenomenological theory of the non-linear electron magnetotransport, accounting the pair electron correlations
within a classical-mechanics picture (that is, the memory effects in the inter-particle scattering), was constructed.

The pair correlations  are  induced by subsequent (``extended'') collisions  of some electrons joined   in pairs due to the cyclotron rotation
\{see  Fig.~1(a,b) and Ref.~\cite{new}\}.  The probability for an electron~1 to pass a cyclotron circle  without collisions
with ``third'' electron~3 (in order to form an ``extended'' collision   with a given electron~2) can be  estimated as: $   P (B) =e^{-T/\tau_q} $,
 where  $\tau_q$ is some inter-particle  scattering time and  $ T=2\pi/\omega_c $ is the cyclotron period. In an almost ideal Fermi gas $\tau_q$
is the electron-electron departure time, but strong pair correlation may substantially change the essence and magnitude of $\tau_q$.

 The pair correlations also lead to the memory  effects in the hydrodynamic-like motion equations~\cite{new},  that is to
the dependence of the fluid dynamics  in the current moment $t$  on the fluid variables in  the proceeding moments  $t'=t-N_r T$,
corresponding to $N_r$ cyclotron periods ago~\cite{new}.  Herewith we do not study the regimes of too high magnetic fields
when high-order memory effects, related to many recollisions, $N_r \gg 1$,  appears.  Apparently, the regime with only few recollisions,
 $N_r \sim 1$,  takes place in relatively low magnetic fields, $ \omega_c \lesssim 2\pi/\tau_q $~\cite{new}.

Here we consider the slow varying flows, with the frequencies $\omega$ much smaller that the shear stress relaxation rate~$1/\tau_2$,
within the model of Ref.~\cite{new}.  We study the simple ``quasi one-dimensional'' flows in long relatively narrow samples with
the width $W$ much smaller  than the plasmon wavelength \{see Fig.~1(a) and Refs.~\cite{vis_res_2,new}\}.  For this geometry,
the $y$-component of  the flow velocity, $V_y$, related  to an internal electric  field $\mathbf{E}_y ^{int} (y,t) $ and
a non-equilibrium charge density~$ e \,\delta n (y,t)$,  is suppressed~\cite{vis_res_2}.  As a result, the motion equations
for the hydrodynamic velocity $ V(y,t)  \equiv V_x (y,t) $ and the shear stress~$  \hat{\sigma} (y,t)  = -\hat{\Pi} (y,t) $ take the form~\cite{ft}:
  \begin{equation}
\label{main_eq_gen}
\left\{
\begin{array}{l}
\displaystyle
    m \: \frac{   \partial V   }{ \partial t   } \, = \,
          e   E \,  - \, \frac{  \partial \Pi_{xy}  }{ \partial y}
 \:,
\\
\displaystyle
 2 \omega_c  \Pi_{xy}   - \frac{ \Pi_{xx} }{ \tau_{2}} -
         \Gamma [V] \:  \hat{\mathcal{T}}_T[\Pi_{xy} ]     =  0
          \: ,
\\
\displaystyle
 - \frac{ \Pi_{xy} }{ \tau_{2} }  - 2 \omega_c  \Pi_{xx} -    \Gamma [V]    \: \hat{\mathcal{T}}_T[\Pi_{xy} ]    =
    \frac{   m \eta_0 }{ \tau_2 }   \: \frac{  \partial V }{ \partial{y} }  \, .
\end{array}
\right.
\end{equation}
where $\hat{\mathcal{T}}_T[\Pi_{ij} ] (y,t)= \Pi_{ij}(y, \, t-T ) $; the electric field  $E \equiv E_x  $ is due  to the applied voltage;
 $e$ and $m$ are the electron charge and the effective mass;  $\tau_{2}$~is  the shear stress relaxation time without
 the memory effects (it was calculated in Refs.~\cite{Novikov,Polini,Alekseev_Dmitriev});
$ \eta_0 =  (v_F^{\eta})^2\tau_2/4 $ is  the viscosity  at zero magnetic field, $v^{\eta}_F $ is the parameter renormalized
by the Landau interaction parameters $F_1$ and $F_2$ as compared to the Fermi velocity~$v_F$~\cite{Semiconductors}.

The extended collisions of electrons leads to the appearance of the ``memory'' non-linear terms of the two types in the fluid motion equations~\cite{new}.  The first ones are the  retarded   contribution  in the relaxation operator  of $\hat{\Pi}$.  This terms are partly
similar to the retarded relaxation terms due to the  extended collisions of non-interacting  electrons with  localized  defects, introduced
in Ref.~\cite{Beltukov_Dyakonov}. The second ones are the ``elastic'' retarded effect in the inter-particle interaction which is expressed via
 the perturbations of the Landau parameter~$F_{2}$ in an inhomogeneous flow.  The second type terms  enter in the second and
 third equations for $\hat{\Pi}$ in Eq.~(\ref{main_eq_gen}) for the case of fast flows when the derivatives $\partial \Pi_{ik} / \partial t$
become comparable with the other terms in those equations.  As  $\omega \tau_2 \ll 1 $,  we omit the time derivatives  of $\hat{\Pi}$ and
the retarded terms of second type disappear.

In Eqs.~(\ref{main_eq_gen}) the terms $ - \Gamma [V] \,    (\Pi_{kl})|_{y,t-T}$ in Eq.~(\ref{main_eq_gen}) are the memory terms of the first
type:  they describe the retarded relaxation  of $\hat{\Pi}$ due to the  extended collisions with one return  [see Fig.~1($b$)]. Such
 extended collisions are sensitive to the macroscopic motion of the fluid in the precedent moments, $t'<t$, via the forces acting on quasiparticles, induced by the inhomogeneous perturbations of the quasiparticle energy spectrum and other effects~\cite{new}.
  The retarded relaxation coefficient   $ \Gamma [V]  $  within such picture depends mainly on the shift (``mismatch'') $  \Delta_{xy} (y,t)
= \varepsilon _{xy} (y,t) -  \varepsilon _{xy} (y, t-T) $ of the  $xy$-strain tensor  component in one cyclotron period \{see Fig.~1(b)
and Ref.~\cite{new}\}. In short,   the parameter controlling  the difference of the inter-particle scattering cross sections at
the current moment~$t$ and at the previous  scattering moment, $t'=t-T$, is the ratio $\Delta r /a_B$, where $\Delta r \sim \Delta_{xy}  |\mathbf{r}_1-\mathbf{r}_2|$  is the characteristic difference  of the impact parameters with and without  accounting   of  the fluid motion
on the electron trajectories [shown in Fig.~1(b) by blue arrows], $\mathbf{r}_{1,2}$ are the centers of trajectories of electrons~$1$ and~$2$,
and $a_B$ is the characteristic (Bohr) radius of the electron screened Coulomb potential.

In considered geometry and at  $\omega \ll 1/\tau_2$,  the value $ \varepsilon_{xy} (y,t) $ characterizes the inelastic displacement
of fluid elements. The shift  of the strain tensor is expressed via the velocity gradient as:  $ \Delta _{xy} (y,t) =
\int _{t-T} ^t dt' \:  \partial V(y,t') /\partial y $.  The mean value of $|\mathbf{r}_1-\mathbf{r}_2|$ in $\Delta r$ is estimated as
the cyclotron radius~$R_c= v_F/\omega_c$. For a small-amplitude flow,  the retarded relaxation coefficient~$ \Gamma [V] (y,t) $
is expanded into power series  by~$ \Delta r /a_B $ and, thus, by~$ \Delta_{xy} (y,t) $:
\begin{equation}
\label{Gamma_ex}
\displaystyle
   \Gamma  [ V(y,t')](y,t)   \, = \,  1/\tau_2'  \,-  \,  \alpha  \,  [\Delta_{xy} (y,t)]^2
     \:,
\end{equation}
where  both the values  $ 1/\tau_2'  $ and $ \alpha $ are proportional to the probability $P (B) =e^{-T/\tau_q} $  for a particle
in a pair  to make a cyclotron rotation without collisions with other (``third'') particles. We see that the first term in
 $\Gamma$~(\ref{Gamma_ex}) is the retarded  contribution to the shear stress relaxation rate from the extended collisions  in
an unperturbed fluid, while the second one is the contribution related to the fluid motion in the lowest (second)  order by the flow amplitude.
 In Ref.~\cite{new} the following estimate was obtained for $\alpha$ from a qualitative consideration  of
the extended collision, briefly cited above:
\begin{equation}
\label{alpha}
\displaystyle
    \alpha    \, = \,  C_\alpha  \,    P(\omega_c) \, (R_c/a_B)^2 \, (1/\tau_2)
     \:,
\end{equation}
where  $ C_\alpha  $ is a numeric constant. The rate  $ 1/\tau_2'  $ is estimated  as $C_0 \, P(\omega_c) \,(1 / \tau_2)$ with some other
numeric constant $C_0$. For simplicity, in further  formulas we will replace the total rate (unperturbed by the fluid motion),
$ 1/\tau_2 + 1/\tau_2' $  just by~$1/\tau_2$, and the parameter~$v_F^\eta$ just by~$v_F$.

For slow  flows, $ \omega \ll 1 / \tau_2 $, the value $V(y,t') $ is almost constant in time interval $(t-T,t)$, however the displacement
 $u_x(y,t')$ of the fluid elements is linear by $t'$. Thus for $\Delta_{xy}$ we have: $\Delta_{xy} (y,t) =  T \, \partial V(y,t) /
\partial y  $ and for the total rate of the shear stress relaxation rate, accounting the fluid motion and the memory effects we obtain:
$ 1/\tau_2 \,- \,T_\Delta \,   [\partial V(y,t) /  \partial y]^2 $, where the characteristic time $T_\Delta $ is:  $ T_\Delta  =
 C_\alpha \, P(\omega_c) \, (R_c/a_B)^2 \,T \, [2\pi /  (\omega_c \tau_2)] $.  The last value depends on magnetic field as   $ P(B)/B^4 $.

 \begin{figure}[t!]
\centerline{\includegraphics[width=.99 \linewidth]{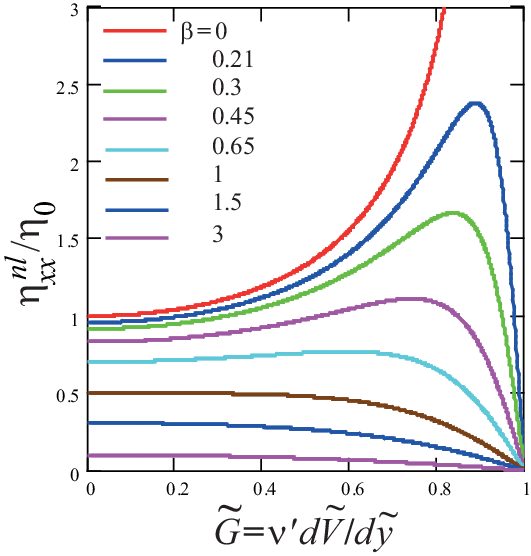}}
\caption{
Dependence ({\em schematic}) of the dimensionless   non-linear viscosity coefficient  $\eta_{xx}^{nl} / \eta_0  $~[see~(\ref{eta_nl})]
 on the dimensionless  velocity gradient~$ \widetilde{G} = \nu' \, \partial \widetilde{V} /\partial \widetilde{y} $ at different
dimensionless magnetic fields $\beta$.  It is seen that, depending of magnetic field, the non-linear viscosity increases or decreases
with the magnitude of the flow gradient~$ \widetilde{G} $, or even can be non-monotonic with~$ \widetilde{G} $.
}
\end{figure}

 The second and the third lines in Eq.~(\ref{main_eq_gen}) yield to the following  relation between $\Pi_{xy}(y,t)$
and~$\partial V (y,t)/ \partial y  $:  $  \:   \Pi_{xy}  \, =\,    - \, m\, \eta _{xx}^{nl}  [V] \: \partial V/ \partial y  \: , $
where we have introduced the  non-linear viscosity coefficient~$ \eta _{xx}^{nl}  [V] (y,t)$:
\begin{equation}
\label{eta_nl}
 \eta _{xx}^{nl}  [V]  = \frac{\eta_0 }{ \displaystyle
   g +
    \displaystyle  (2\omega_c\tau_2) ^2 /g }
   ,
   \quad
   g = 1 - \tau_2  T_\Delta  \Big( \frac{\partial V }{ \partial y} \Big) ^2
   \,.
\end{equation}
The resulting motion  equation for slow flows is~\cite{f2}:
\begin{equation}
\label{motion_eq_fin}
   \frac{ \partial V }{ \partial t }
    \, = \,
   \frac{eE}{m}
   \,   + \,
   \frac{\partial}{\partial y} \Big( \, \eta _{xx}^{nl}  [V] \,
     \frac{ \partial V}{ \partial y } \, \Big)
     \, .
\end{equation}
Formula (\ref{eta_nl}) establishes  that the memory effects lead to the dependence of viscosity on the velocity gradient. In other words,
 the electron fluid becomes non-Newtonian, dilatant at classically low magnetic fields ($\eta_{xx}^{nl}$ increases with
$ G=\partial V/\partial  y $ at $ \omega_c \tau_2 <1 $) and pseudo-plastic at high magnetic fields ($\eta_{xx}^{nl}$ decreases
with $G $ at $\omega_c \tau_2 > 1 $). Such behavior of $\eta_{xx}^{nl}(G)$  is illustrated in Fig.~2. Note that  at $\omega_c \tau_2 <1 $,
 and at sufficiently large $G$ the viscosity become non-monotonic with $G$. Such case can be considered as a ``extreme'' type
of non-Newtonian fluids.

{\em 3. Stationary flow in long sample. }
We consider a stationary Poiseuille flow   in a defectless long sample  as a minimal model to study magnetotransport in the 2D electron
fluid in non-linear regime.

Let us introduce the dimensionless variables $\widetilde{y} = y/W$ and $\widetilde{V} = V/ V_0$,  where $V_0 = eE W^2 / (m\eta_0) $ is
the characteristic magnitude of the hydrodynamic velocity in the stationary Poiseuille flow   in the linear by $E $~regime
in the absence of magnetic field.  Accounting the symmetry of~$V(y)$ with respect to the sample center, $y=0$,
equation~(\ref{motion_eq_fin})  takes the form:
\begin{equation}
\label{motion_eq_stat}
   \frac{ \displaystyle  d \widetilde{V} / d \widetilde{y} }
   { \displaystyle
  g +
   \beta ^2/ g }
   =
   - \widetilde{y}
   \, ,
   \quad \;\;
    g =   1 - \nu' \,   \Big( \frac{d \widetilde{V} }{ d\widetilde{y}}  \Big) ^2
  \, ,
\end{equation}
where $\beta = 2\omega_c\tau_2 $,  $\nu' (\beta)  \,= \,  C_\nu \, f(\beta) \, \nu $ is the dimensionless parameter of non-linearity,
\begin{equation}
\label{f_,_nu}
  f(\beta)  =  \frac{ P ( B ) }{ \beta^4} = \frac{ e^{-r/\beta}}{ \beta^4}
 \, , \quad \;\;
  \nu  =   \Big( \, \frac{ l_2 }{a_B} \frac{ eEW }{ m v_F^2 } \, \Big)^2
 \,,
\end{equation}
$ r = 4\pi\tau_2/\tau_q $, $l_2 = v_F \tau_2$, $ \nu    $ is the magnetic-field-independent parameter  of non-linearity,  and~$ C_\nu $
is a numeric coefficient related to  $ C_\alpha $ in Eq.~(\ref{alpha}) as: $ C_\nu =C_\alpha (2\pi)^2 \,2^4\,4^2 $.

One should impose some boundary condition on the velocity $V(y)$ at the sample edges, $y=\pm W/2$. For simplicity, we chose the sticking
boundary conditions corresponding to the rough edges: $ V|_{y=\pm W/2} = 0 $.

Equation~(\ref{motion_eq_stat}) is the ordinary differential equation of the first order of the special type: the equation not containing
the very unknown function $\widetilde{V}(\widetilde{y})$, but containing only $ d\widetilde{V} / d\widetilde{y} $ and  $\widetilde{y}$.
 Its solution is solved analytically  and is expressed  in the parametric form:
\begin{equation}
 \label{solution}
  \begin{array}{l}
  \displaystyle
  \widetilde{y}(u) = - \frac{ (1- \nu'u^2) \, u  }{ \beta^2 + (1 - \nu'u^2)^2 }
  \:,
  \:\;\;
  \widetilde{V} (u) = \widetilde{ V} _1 - \frac{Y(u)  }{\nu'}\:,
  \\
  \\
  \displaystyle
Y(u)  =
    \frac{ \beta^2 +  1- \nu'u^2   }{ \beta^2 + (1- \nu'u^2) ^2 }
    +
    \frac{\ln[  \beta^2 + (1 - \nu'u^2) ^2  ] }{4  }
    \,.
 \end{array}
\end{equation}
where $u $ is the parameter which varies in the interval $ (-u_0,u_0) $, whose edges corresponds to the sample edges:
$ \widetilde{y}( \pm u_0) = \pm  1/2 $,   and  $ \widetilde{V}_1 $ is the constant that satisfies the boundary
conditions~$\widetilde{V}( \pm u_0 ) =0$.   The equation for the boundary points $\pm u_0$,
$ \widetilde{y}( u_0) = 1/2 $, is the fourth order algebraic equation. Its proper (being real and minimal by the absolute value)
 solution $u_0$ is given by well-known Ferrary's formula.

The flow profile calculated  by Eq.~(\ref{solution}) is shown in Fig.~1(c). It is seen that accounting of the dependence of the viscosity
on the velocity gradient  leads only to small deviation of the shape of the profile   from the parabolic one corresponding to
the linear Poiseuille flow: the profile of the flow of non-Newtonian fluid becomes ``more triangular''.

{\em 4. Results and  discussion.  }
By use of a numerical analysis of the obtained analytical solution~(\ref{solution}), we have determined  the critical values of
the non-linearity parameters  $ \nu'_{cr}(\beta) $ and $ \nu _{cr}(\beta) = \nu'_{cr}(\beta)  /[C_\nu  f(\beta) ]$. Namely,
 below  $ \nu '_{cr} (\beta) $,  $ \nu ' < \nu '_{cr} (\beta) $, the functions $u(y)$  and $V(y)$ remails well-defined (unambiguous)
 and the developed model leads to a smooth flow $V(y)$.  The functions $ \nu '_{cr} (\beta) $ and $ \nu _{cr} (\beta) $
 are monotonic, namely, decreases with $\beta$ from very large values much, $ \nu ' _ {cr} , \, \nu _{cr} \gg 1 $,
at $\beta \ll 1 $  up to  very small values, $ \nu ' _ {cr} , \, \nu _{cr} \ll 1 $, at $ \beta \gg 1 $ (see Fig.~S1 in SM~\cite{SM}).

 At higher values of   the non-linearity parameter, $ \nu >\nu _{cr}   $,  the function $\widetilde{y}(u)$ in Eq.~(\ref{solution})
becomes non-monotonic, that can correspond to unstable space-inhomogeneous   solutions.  The last ones can be found only
within some more general model, accounting additional effects those stabilize the flow shape  (for example, the effects of
the sample edges, other type of relaxation processes and non-linearities, and so on).

The total electric current  can be presented as:  $I=I_0 \, \widetilde{I}$, where  $ I_0 = e^2 n_0 E W^3 / (m \eta_0 ) $ is its amplitude
in the linear regime, and $\widetilde{I} = \int_{-W/2} ^{W/2} d \widetilde{y} \: \widetilde{V}( \widetilde{y} )$.  It is of interest
to find the differential resistance  $R=dE/dI = 1/(dI/dE) $, $R=R(\beta,I)$, being the value which exhibits the features
of  the dependence $I(B,E)$ more clearly. From the form of the factor~$ I_0$  and  the nonlinearity parameter~$\nu (E) $, we obtain:
$  R (E)^{-1}\, = \, dI/dE \,  =\, (I_0/E) \, (\,\widetilde{I}    \,+\ 2\nu \, d \widetilde{I} / d \nu \, )   \:. $
Using this formula and Eqs.~(\ref{solution}), we calculate the current $I$, the resistance $\varrho = E/I$, and the differential
 resistance  $R$ as functions of the parameters~$\beta$ at different~$\nu$.

In Fig.~3 we present  the dimensionless values  $ \tilde{\varrho }=1/\widetilde{I} =1/(I/I_0) $  and  $ \widetilde{R} = (I_0/E) / (dI/dE) $
as functions of the parameter $\beta \sim B$  for different   $\nu$.  We see a non-monotonous  behavior of $\varrho(B)$ and $R(B)$  at smaller
magnetic fields, $\beta <1$: the appearance of  more or less distinct   maximum at some $\beta^*$ with the amplitude growing with $\nu$.
At magnetic fields  $\beta \gtrsim 1 $ the value $R(\beta)$  decreases with  $\nu$.  At some $\beta_c$ a break of curves  $R(\beta)$
is observed (see Fig.~3). It  originates from arising of the uniqueness  in the function $\widetilde{y}(u)$ at  $\nu \to \nu_{cr}(\beta)$,
that is, to the reaching of a maximal amplitude of the  non-linear flow, when the developed theory does not lead
to unambiguous results and  the system can become unstable and inhomoheneous. We have performed the calculations for  the limiting case~$\tau_q\sim\tau_2$, namely~$r=3$ (and~$r=7$, see SM). Exactly for this case,  the non-linear features  in $\varrho(B)$
and $R(B)$ within the considered model are strong.

Calculations for higher values of $\beta$  and $r$ shows  that,  with the increase of~$r$, the diapason  of a substantial deviation
of $R(\beta)$ from $\varrho (\beta)$ moves to higher values of~$\beta$ (see Fig.~S2 in SM). Herewith the magnitudes of the non-linearity
 features in $R(\beta)$ becomes smaller. The break of the curves with the increase of $\beta$ is observed at higher~$\beta_c$.

In this way, the bright non-linear behavior of $R(\beta)$ (see Fig.~3) takes place only at relatively small $r$  corresponding to $\tau_2
\sim \tau_q$,   whereas for conventional Fermi systems (Fermi  gas and Fermi liquid) the kinetic theory  yields: $\tau_{ee,2}
 \gg \tau_{ee,q}$~\cite{Novikov,Alekseev_Dmitriev}. One of the reason of the unusual relation between $\tau_{2}$ and $ \tau_{q}$
in our theory can be related to the strong interparticle correlation consisting in formation of electron, those suffer several
re-collisions (see Fig.~1). The relation between the parameters $\tau_{2}$ and $ \tau_{q}$
in model~\cite{new} and, in particular,  in the current model can be very different  from the relation between
the single-electron relaxation times~$\tau_{ee,2}$ and $ \tau_{ee,q}$ in a conventional Fermi-systems without memory effects. This issue
needs further microscopic consideration.

  \begin{figure}[t!]
\centerline{\includegraphics[width=.99 \linewidth]{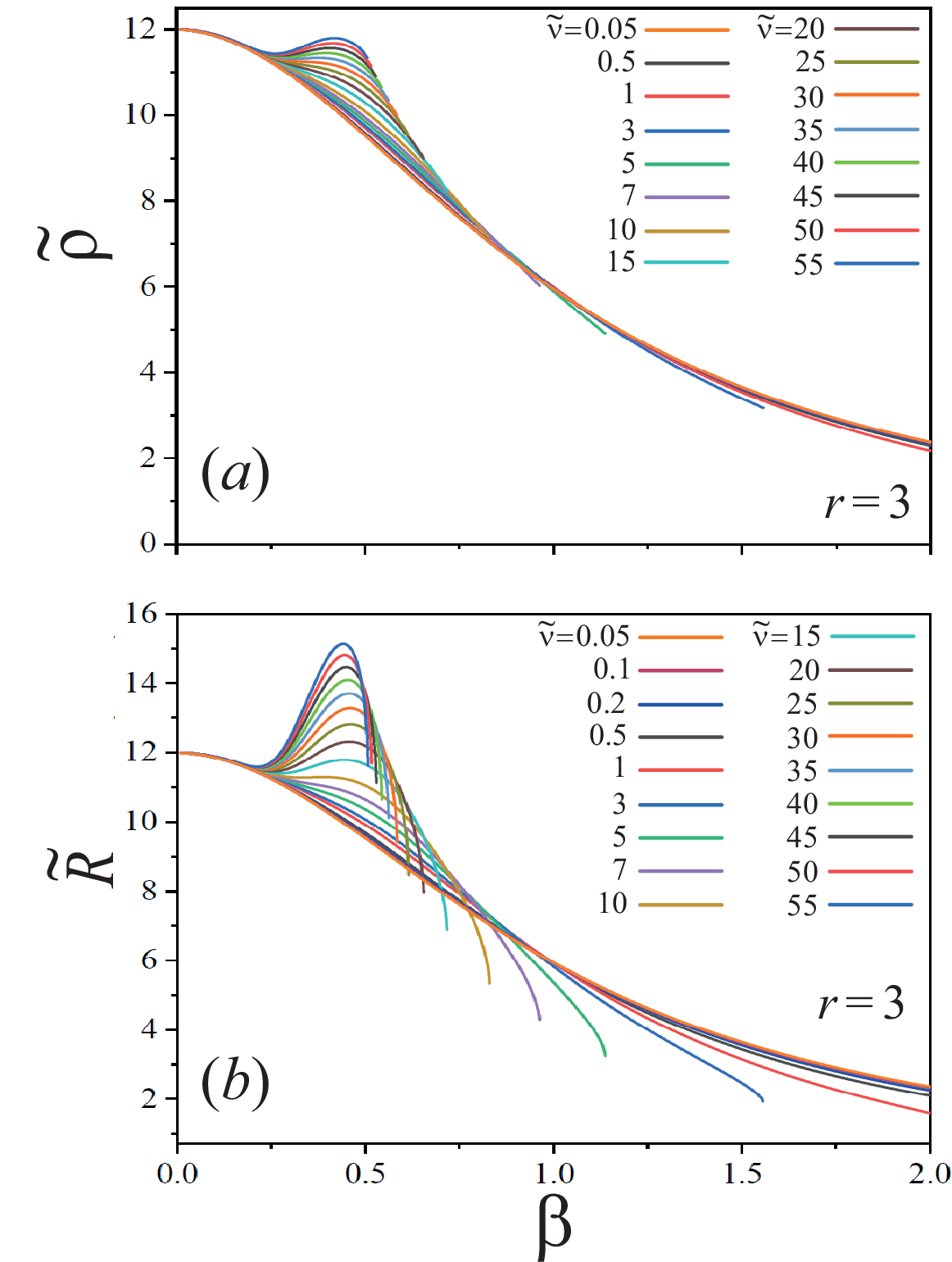}}
\caption{
($a$,$b$): Dimensionless resistance~$ \widetilde{  \varrho }= 1/\widetilde{I} = 1/(I/I_0) $~($a$)  and dimensionless differential
resistance~$ \widetilde{ R } \, = \, (I_0/E_0) \, dE/dI  $~($b$) as functions of dimensionless magnetic field~$\beta$ for different
nonlinearity parameters $ \widetilde{\nu} \, = \,   C_ \nu \, \nu  $. The calculation is performed for   $r=3$, that corresponds to
the limiting magnitude of the ratio $\tau_q/\tau_2 \sim 1 $.
  }
\end{figure}

In SM~\cite{SM} we compare the experimental results of Ref.~\cite{non-lin_hydr_1}, where the non-linear magnetrotransport of high-quality
GaAs quantum wells was examined,  with the predictions of our theory. We also discuss in~\cite{SM} some properties of the samples, studied in~\cite{non-lin_hydr_1}, those seem to be important. To summarize this comparison,
the appearance of a maximum in the differential magnetoresistance~$R(B)$ at a small magnetic field $B^*>0$ was observed
in experiment~\cite{non-lin_hydr_1}  and is predicted by the theory [see Fig.~3(b)]. So a part of  the experimental properties
of $R(B,I)$ can be explained by the proposed here  mechanism of the non-linear magnetotransport due to the pair correlations.

It is of interest to study other non-linear effects in the hydrodynamic  transport, first of all,  the heating of electrons   by the current.
 Our preliminary estimations shows that   some other propertied of the non-linear magnetoresistance  observed in experiment~\cite{non-lin_hydr_1}
(the smooth dependencies of $R$ on $I$ at $B=0$ and at $B\gg B^*$), as well as for the non-linear phenomena observed in
experiments ~\cite{non-lin_hydr_3,non-lin_hydr_4},  can be explained by the heating mechanism.

{\em 5. Conclusion. }
We have developed a theoretical model for highly non-linear hydrodynamic magnetotransport in a 2D electron fluid.  The pair correlations
 due to the ``extended collisions''  of paired electrons on cyclotron orbits   lead to the non-linear memory effect in the relaxation
of the shear stress. For slow flows, the last effect induces   the dependence of the electron viscosity on the hydrodynamic velocity gradient,
that  is to the non-Newtonian behavior of the electron fluid, whose type (pseudoplastic, dilatant, or more a complex type)  is controlled
by  magnetic  field.   We have compared our results with experiment~\cite{non-lin_hydr_1} on non-linear dc magnetotransport in
the high-quality GaAs quantum wells.  The proposed non-linearity mechanism can explain, at least,  a part  of the properties of
 the observed non-linear magnetoresistance.

We thank M. I. Dyakonov for attracting our attention to the experimental  results on non-linear magnetotransport (first of all, results
 of Ref.~\cite{non-lin_hydr_1}),  on explanation of which is aimed this work, as well as for fruitful discussions and kind support.
We thank A. N. Afanasiev for attracting our attention to the concepts of non-Newtonian fluids, those lead to the current view
on our results.

A part of this work (analytical calculations: derivation of the non-linear hydrodynamic equation and their analytical solution;
Sections~2 and~3) was financially supported by    the Russian Science Foundation (Grant No.~25-12-00093).
One of us (P.~S.~A.) is grateful to the Foundation for the Advancement of Theoretical Physics and Mathematics  ``BASIS''
(Grant No.~23-1-2-25-1)   for support of this part of the work.
A part of this work (numerical calculations:  numeric solution of the non-linear hydrodynamic equation; comparison of the obtained
 theoretical results   with the experiment;  Section~4)  was carried out within the state assignment of Ministry of Science
and Higher Education  of the Russian Federation.

\clearpage

\setcounter{equation}{0}
\setcounter{figure}{0}

\renewcommand{\thefigure}{S\arabic{figure}}
\renewcommand{\thesection}{S\Roman{section}}
\renewcommand{\theequation}{S\arabic{equation}}

\setcounter{equation}{0}
\setcounter{figure}{0}

\renewcommand{\thefigure}{S\arabic{figure}}
\renewcommand{\thesection}{S\Roman{section}}
\renewcommand{\theequation}{S\arabic{equation}}

\onecolumngrid
\begin{center}

{\Large  {\bf Supplemental material   to the article
 ``Analytical model for non-linear magnetotransport  in viscous  electron fluid'' }
\linebreak
}

{
\large P. S. Alekseev and M. A. Semina
\linebreak \linebreak}
{\small
 Ioffe  Institute, Politekhnicheskaya 26,
  194021,   St.~Petersburg,   Russia
\linebreak
}
\end{center}

{\small
Here we present
 the details of our theoretical results,
 compare our results with experimental data on non-linear magnetotransport in
high-purity GaAs quantum wells, 
discuss the properties of the samples which can be important for the realization of
 the considered non-linear hydrodynamic regime, and discuss other possible mechanisms of non-linearity 
  in these systems.
\linebreak
\linebreak
\linebreak}
\twocolumngrid

{\em 1. Expressions and properties of constructed analytical solution.}
The Navier-Stokes-like equation for the slow flows (with the frequencies $\omega \ll  1/\tau_2$)  in a long sample takes
the form (see Sections~2 and~3 in the main text):
\begin{equation}
 \label{motion_eq_fin}
   \frac{ \partial V }{ \partial t }
    \, = \,
   \frac{eE}{m}
   \,   + \,
   \frac{\partial}{\partial y} \Big( \, \eta _{xx}^{nl}  [V] \,
     \frac{ \partial V}{ \partial y } \, \Big)
     \, ,
\end{equation}
where $\eta _{xx}^{nl}  $ is the non-linear viscosity:
\begin{equation}
\label{eta_nl}
 \eta _{xx}^{nl}  [V]  = \frac{\eta_0 }{ \displaystyle
   g +
    \displaystyle  (2\omega_c\tau_2) ^2 /g }
   ,
   \quad
   g = 1 - \tau_2  T_\Delta  \Big( \frac{\partial V }{ \partial y} \Big) ^2
   \,,
\end{equation}
$T_\Delta = 2\pi C_\alpha P(\omega_c) (R_c/a_B)^2 T /(\omega_c \tau_2)$;  $C_\alpha$ in the numeric constat entering the memory term
 in the shear stress relaxation terms in equations for $\partial \hat{ \Pi } / \partial  t $ [see Eqs.~(2) and~(3) in the main text];
$\tau_2$ is the full shear stress relaxation time; $R_c$, $\omega_c$, and $T$ are the cyclotron radius, the cyclotron frequency,
and the cyclotron period; $\eta_0$ is the electron viscosity in zero magnetic field; $P(\omega_c) = e^{-2 \pi / (\omega_c \tau_q)}$ is
the probability for an electron to make a full cyclotron rotation without collisions with other electrons; $\tau_q$ is the characteristic
scattering time  which is equal   to scattering departure time in the Fermi systems without the correlations due to the inter-particle
interaction and can substantially changed at strong inter-particle correlations  (see  discussion in the main text and below); $a_B$ is
the characteristic radius of the inter-particle interaction (approximately  the Bohr radius).

We consider a stationary Poiseuille flow   in a defectless long sample  with rough edges as a minimal model to study magnetotransport in
the 2D electron fluid in non-linear regime. The simplest sticking boundary conditions corresponding to the rough edges, $V|_{y=\pm W/2}=0$,
are used. We introduce the dimensionless variables $\widetilde{y} = y/W$ and $\widetilde{V} = V/ V_0$,  where $V_0 = eE W^2 / (m\eta_0) $
 is the characteristic magnitude of the hydrodynamic velocity in the stationary Poiseuille flow   in the linear by $E $~regime.
Accounting the symmetry of~$V(y)$ with respect to the sample center, $y=0$, equation~(\ref{motion_eq_fin})takes the form:
\begin{equation}
\label{motion_eq_stat}
   \frac{ \displaystyle  d \widetilde{V} / d \widetilde{y} }
   { \displaystyle
  g +
   \beta ^2/ g }
   =
   - \widetilde{y}
   \, ,
   \quad \;\;
    g =   1 - \nu' \,   \Big( \frac{d \widetilde{V} }{ d\widetilde{y}}  \Big) ^2
  \, ,
\end{equation}
where $\beta = 2\omega_c\tau_2 $,  $\nu' (\beta)  \,= \,  C_\nu \, f(\beta) \, \nu $ is the dimensionless parameter of non-linearity,
\begin{equation}
\label{f_,_nu}
  f(\beta)  =  \frac{ P ( B ) }{ \beta^4} = \frac{ e^{-r/\beta}}{ \beta^4}
 \, , \quad \;\;
  \nu  =   \Big( \, \frac{ l_2 }{a_B} \frac{ eEW }{ m v_F^2 } \, \Big)^2
 \,,
\end{equation}
$ r = 4\pi\tau_2/\tau_q $, $ \nu    $ is the magnetic-field-independent parameter  of non-linearity, $l_2 = v_F \tau_2$ is   the shear
stress  scattering length, and $ C_\nu $ is a numeric coefficient related to the coefficient in formula~(3) in the main text
 as:~$ C_\nu = 2^{10}\pi^2 C_\alpha   $.

Equation~(\ref{motion_eq_stat}) is the ordinary differential equation of the first order of the special type: the equation not containing
the very unknown function $\widetilde{V}(\widetilde{y})$, but containing only the derivative $ d\widetilde{V} / d\widetilde{y} $
 and the variable $\widetilde{y}$. Its solution is expressed  in the parametric form by quadratures.   After analytical calculations
of integrals, we obtain the result:
\begin{equation}
 \label{solution}
  \begin{array}{l}
  \displaystyle
  \widetilde{y}(u) = - \frac{ (1- \nu'u^2) \, u  }{ \beta^2 + (1 - \nu'u^2)^2 }
  \:,
  \:\;\;
  \widetilde{V} (u) = \widetilde{ V} _1 - \frac{Y(u)  }{\nu'}\:,
  \\
  \\
  \displaystyle
Y(u)  =
    \frac{ \beta^2 +  1- \nu'u^2   }{ \beta^2 + (1- \nu'u^2) ^2 }
    +
    \frac{\ln[  \beta^2 + (1 - \nu'u^2) ^2  ] }{4  }
    .
 \end{array}
\end{equation}
where $u $ is the parameter which varies in the certain interval $ (-u_0,u_0) $, whose edges corresponds to the sample edges:
$ \widetilde{y}( \pm u_0) = \pm  1/2 $,   and  $ \widetilde{V}_1 $ is the constant which corresponds to the boundary
conditions~$\widetilde{V}( \pm u_0 ) =0$.   In this way, the boundary points $\pm u_0$ are determined from the equation
$ \widetilde{y}( u_0) = 1/2 $, which is the fourth order algebraic equation. Its proper (being real and minimal by the absolute value)
 solution $u_0$ is given by well-known Ferrary's formula.

  \begin{figure}[t!]
\centerline{\includegraphics[width=.99 \linewidth]{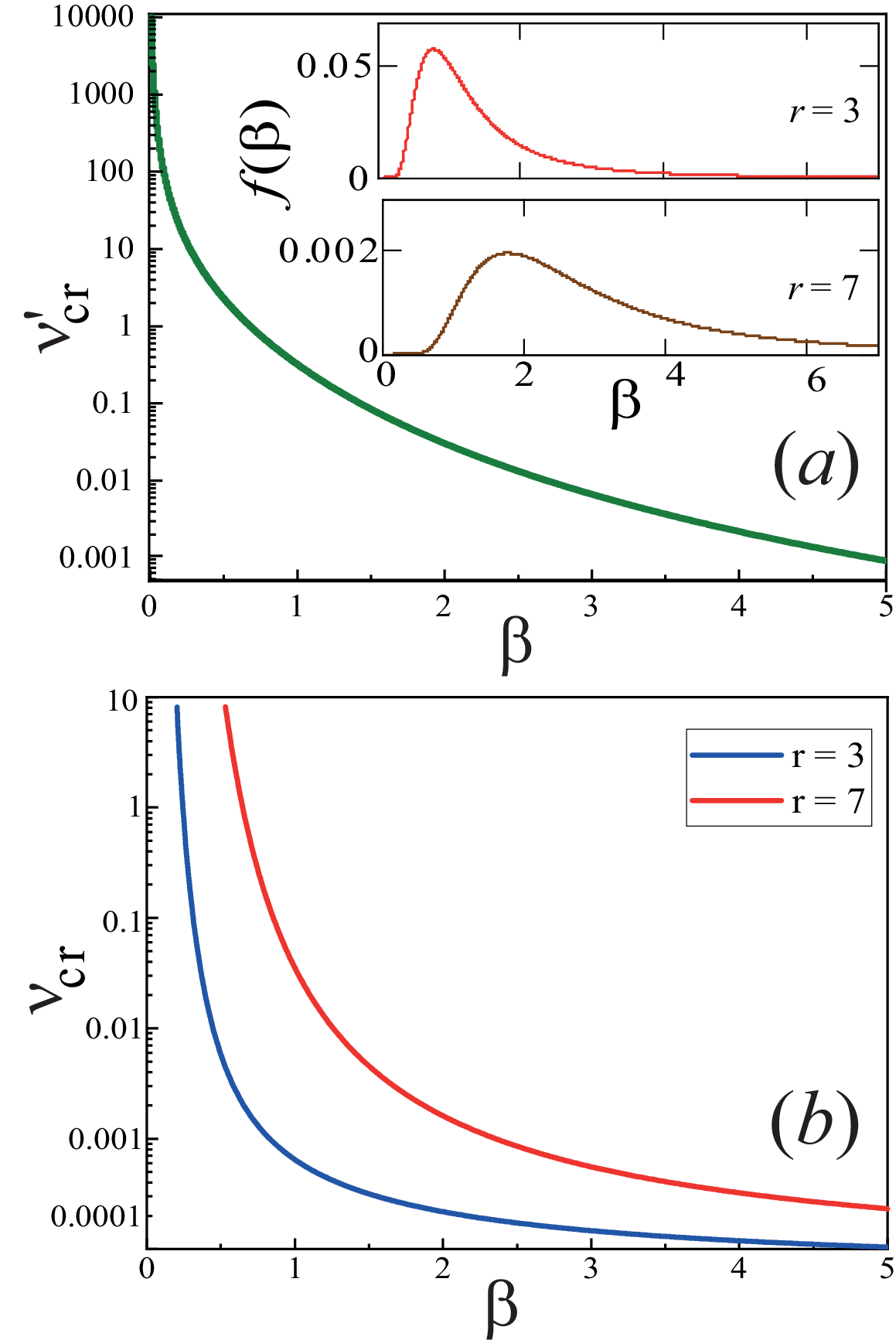}}
\caption{
 Critical values of the parameters  $\nu'(\beta) $~[panel~($a$)] and $ \nu (\beta) \sim \nu(\beta) / [2^{10} f(\beta) ]  $~[panel~($b$)]
 corresponding to maximal possible amplitude of the non-linear stable flows within the current model.  Note that here we omit
the numeric factor $C_\alpha$  in Eq.~(3) in the main text,  implying  $C_\alpha \sim 1$;   this factor is not established up now
[see the discussion in Section~2 of the main text]. Inset: the magnetic-field-dependent factor $ f (\beta) =e^{-r/\beta}/\beta^4 $,
which connects the non-linearity    coefficients~$\nu'(\beta) = f (\beta) \, \widetilde{\nu}$  and~$\widetilde{\nu} = C_\nu \, \nu$
for the cases of the two small values  of the parameter   $ r = 4 \pi \tau_2 /\tau_q$, $r=3,\,7$.
   }
\end{figure}

By use of a numerical analysis of the obtained analytical solution~(\ref{solution}), we have determined  the critical values of
the non-linearity parameters  $ \nu'(\beta) $ and $ \nu (\beta) = \nu'(\beta)  /[C_\nu  f(\beta) ]$. Namely, below  $ \nu '_{cr} (\beta) $,
 $ \nu ' < \nu '_{cr} (\beta) $, the functions $u(y)$  and $V(y)$ remails well-defined (unambiguous) and the developed model leads
to a smooth flow.   The result is presented at Fig.~S1.    The parameters $ \nu '_{cr} (\beta) $ and $ \nu _{cr} (\beta) $
 are the monotonic functions, decreasing with $\beta$ from very large values much, $ \nu ' _ {cr} , \, \nu _{cr} \gg 1 $,
at $\beta \ll 1 $  up to  very small values, $ \nu ' _ {cr} , \, \nu _{cr} \ll 1 $, at $ \beta \gg 1 $.

At higher values of   the non-linearity parameters~$ \nu $ and~$\nu '  $ ($ \nu > \nu _{cr} , \:\nu ' > \nu ' _{cr}   $) the function
 $\widetilde{y}(u)$ in Eq.~(\ref{solution}) becomes non-monotonic, that apparently corresponds to unstable space-inhomogeneous   solutions.
The last ones can be found only within some more general model with additional important effects those stabilize the flow shape  (for example,
the effects of  the sample edges, other type of relaxation processes and non-linearities, and so on).

The total electric current via the sample is calculated by usual formula: $ I (E)=en_0\int_{-W/2} ^{W/2} dy \: V(y)$. It can be
presented as:  $I=I_0 \, \widetilde{I}$, where  $ I_0 = e^2 n_0 E W^3 / (m \eta_0 ) $ is the amplitude of the current in the linear
regime, and $\widetilde{I} = \widetilde{I} (\nu,\beta) = \int_{-W/2} ^{W/2} d \widetilde{y} \: \widetilde{V}( \widetilde{y} )$
is the dimensionless current depending on the nonlinearity parameter~$\nu \sim E^2 $~(\ref{f_,_nu}) and the dimensionless
magnetic field~$ \beta $.

It is of interest to calculate the differential resistance  $R=dE/dI = 1/(dI/dE) $, $R=R(\beta,I)$, being the value which exhibits features
of  the dependence $I(B,E)$ more clearly and, apparently by this reason, was presented in experimental publication~\cite{non-lin_hydr_1}.
 Here we imply that the total current and the electric field at given~$\nu$ and~$\beta$ are the single-valued functions each of the other,
$ I(E) \leftrightarrow E(I)$. In view of the form of the factor~$ I_0$  and  the nonlinearity parameter~$\nu (E) $, we obtain:
$  R (E)^{-1}\, = \, dI/dE \,  =\, (I_0/E) \, (\,\widetilde{I}   \,+\, 2\nu \, d \widetilde{I} / d \nu \, )   \:. $ Using this formula
and Eqs.~(\ref{solution}), we calculate the current $I$, the resistance $\varrho = E/I$, and the differential  resistance  $R$ as functions
of the parameters~$\beta$ at different~$\nu$.

In Fig.~S2 we present  the values  $ \tilde{\varrho }=1/\widetilde{I} =1/(I/I_0) $  and  $ \widetilde{R} = (I_0/E) / (dI/dE) $ as functions
of the parameter $\beta$  for different  values  of $\nu$.  We have performed the calculations for  the limiting case~$\tau_q\sim\tau_2$,
namely~$r=3$ [panels~($a$,$b$)] and~$r=7$ [panels~($c$,$d$)]. Exactly for such case  the non-linear features  in $\varrho(B)$ and $R(B)$
within the considered mechanism are strong.   These features are  a non-monotonous  behavior of $\varrho(B)$ and $R(B)$  at smaller
magnetic fields, $\beta <1$: the appearance of  more or less distinct   maximum at some $\beta^*$ with the amplitude growing with $\nu$.
At magnetic fields  $\beta \gtrsim 1 $ the value $R(\beta)$  decreases with  $\nu$.  At some $\beta_c$ a break of curves  $R(\beta)$
is observed [see Fig.~S2($b$,$d$)]. It  originates from arising of singularities  in the function $\widetilde{y}(u)$
 at  $\nu \to \nu_{cr}(\beta)$,  that is, to the reaching of a maximal amplitude of the  non-linear flow, when the developed theory
does not lead  to unambiguous results and  the system can become unstable and inhomoheneous.

 Calculations for higher values of $\beta$  and $r$ shows [see Fig.~S2(c,d)] that,  with the increase of~$r$, the diapason
 of a substantial deviation of $R(\beta)$ from $\varrho (\beta)$ moves to higher values of~$\beta$. Herewith the magnitudes
of the non-linearity features in $R(\beta)$ becomes smaller. The break of the curves with the increase of $\beta$ is observed
at higher~$\beta_c$.

{\em 2. Important properties of the samples.}
Let us discuss the role of different defects and disorder in the typical structures where the electron hydrodynamics is realized.

As it was mentioned above, in high-quality GaAs quantum wells, the macroscopic ``oval'' defects are typically present~\cite{d1,d1_new}.
 Recently, arrays of similar defects with controlled densities were made artificially in high-quality GaAs quantum wells~\cite{d2}
and graphene~\cite{d2_new}  samples. The oval defects can control the effective width~$W_{eff}$ of the conductive channel, through which the
electron fluid flows~\cite{je_visc,new}. A rigorous  theory of magnetotransport in teh systems with macroscopic disk defects was developed
in Ref.~\cite{disks}. In experiments~\cite{d1_new,d2_new} it was demonstrated that in samples with such defects hydrodynamic as well 

\clearpage

\onecolumngrid
\begin{center}
  \begin{figure}[th!]
\centerline{\includegraphics[width=0.75 \linewidth]{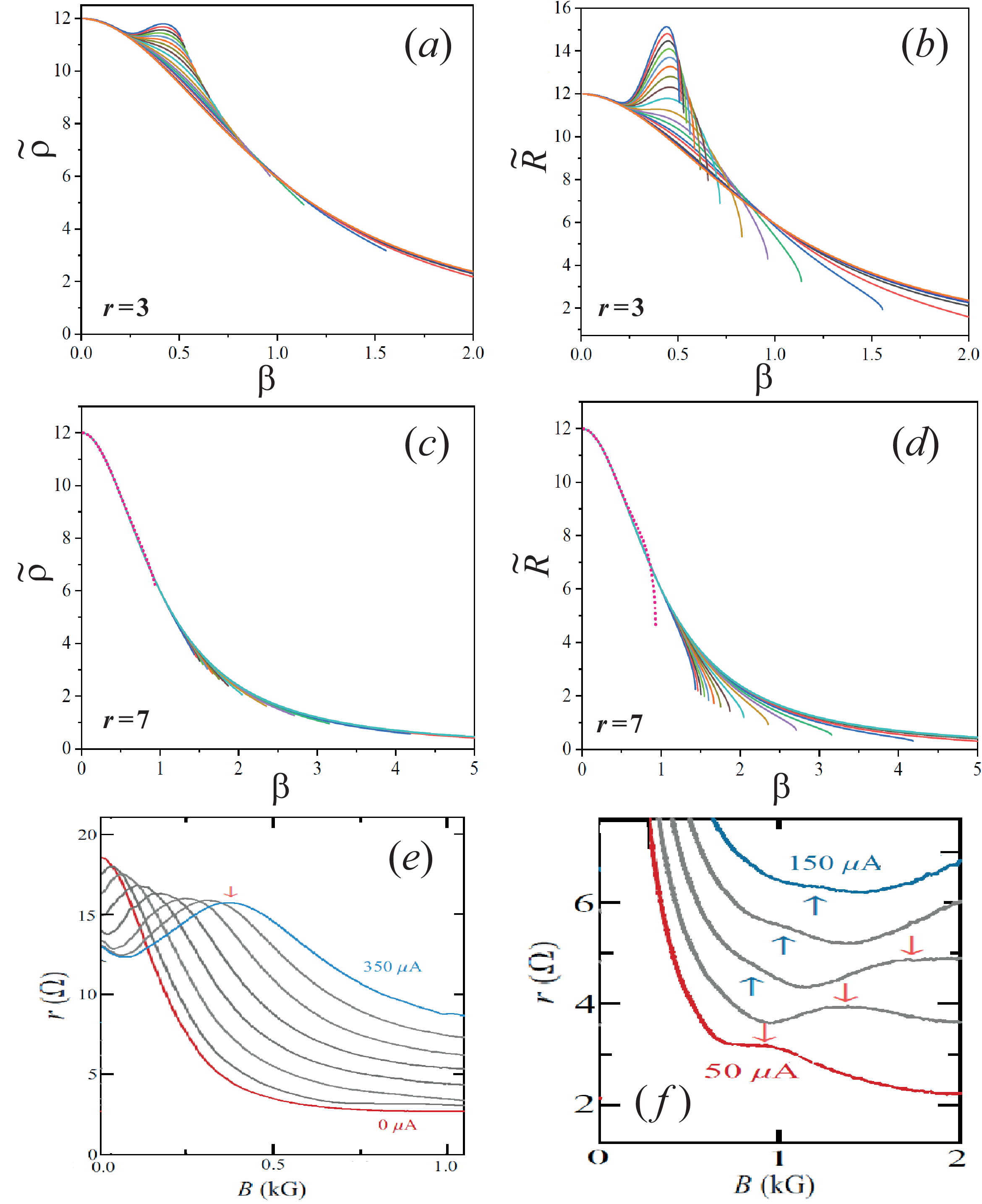}}
\caption{
($a$,$b$):
Dimensionless resistance~$ \widetilde{  \varrho } = \widetilde{E}/\widetilde{I} = (E/E_0)/(I/I_0) $  ($a$) and dimensionless differential
resistance~$ \widetilde{ R } \, = d\widetilde{E}/d\widetilde{I} = \, (I_0/E_0) \, dE/dI $ ($b$) as functions of dimensionless magnetic
field~$\beta$ for different nonlinearity parameters $ \widetilde{\nu} \, = \,   C_ \nu \, \nu   = 55, \, 50, \, 45, \, 40, \, 35, \, 30,
\, 25, \, 20, \,  15,  \, 10, $, $ \, 7,  \, 5, \,  3,  \, 2 , \,  1 , \,  0.5  , \,  0.2 , \,  0.1 , \,  0.05   $ [for the curves
of different colours in the sequence corresponding to the decrease of the amplitude of the deviation of the current curves
$ \widetilde{  \varrho }(\beta,\nu)$, $ \widetilde{ R }(\beta,\nu)$   from the Lorenzian-shape linear approximation curves $  \varrho _0(\beta )
= \varrho (\beta, \nu \to 0 ) = R (\beta, \nu \to 0 ) $]. The calculation is performed for $r=3$, that correspond to the limiting magnitude
of the ratio $\tau_q/\tau_2 \sim 1 $.
($c$,$d$):
The same data as in panels ($a$,$b$) for $r=7$ and a wider diapason of $\beta$, with an additional curves (dotted lines) for a very large
 nonlinearity parameter, $ \nu = 550$.
($e$,$f$):
Experimental data from Ref.~\cite{non-lin_hydr_1} for the differential resistance~$ r =dU/DI$   for ultra-high-quality GaAs quantum wells
at T = 1.5~K for different values of the electric currents $I$ (shown in panels) for a narrower~($e$) and  a wider~($f$) diapason
of magnetic fields, perpendicular to the 2D layer.
 }
\end{figure}

\end{center}
\twocolumngrid

\clearpage

\noindent as superballistic  flows of 2D electrons can be formed.

Between oval defects  and at some distances from the 2D layer, additional  short-range defects also may be present
and affect magnetotransport. The macroscopic defects can lead   a non-uniform distribution of the electron density~$n_0(\mathbf{r})$.
 Such non-uniform electron density distributions over a sample were  directly experimentally observed, for example,
in Ref.~\cite{random_charge_denstiy}. The density $n_0(\mathbf{r})$ also induces a random potential, which is additive with
the short-range disorder potential from microscopic defects between macroscopic ``oval'' defects.

\indent
 Both these two type of the microscopic disorder potential can should lead to  additional scattering of 2D electrons and also change of rates
of inter-particle relaxation rates: $ 1/\tau_{q,2}^{ee} \to 1/\tau_{q,2}^{dis} + 1/\tau_{q,2}^{ee}   $.   In particular, there appears
a temperature-independent  and a temperature-dependent contributions in the shear stress relaxation rate, which control the electron viscosity:
 $1/\tau_2 =1/\tau_2^{ee}(T) + 1/\tau_2^{dis} $~\cite{je_visc,Alekseev_Dmitriev}. The actual presence of these two contribution
 in high-quality GaAs quantum wells were demonstrated by analysis of experimental  data on the giant negative magnetoresistance
at $T \to 0 $~\cite{je_visc,Gusev_1,Alekseev_Dmitriev}.

It is also very possible that defects of different type can substantially affect the memory effects. For example, small-size microscopic defects
can lead to the appearance of  two types of the single electrons and to the additional breaking of the paired electrons by the scattering
on these defects of one of the electrons in a pair. All such processes may be described by some change of the probability
$P(\omega) e^{-2\pi/(\omega_c \tau_q ) }$.  The simplest variant of such change is the adding of some disorder-induced  residual departure rate
in $1/\tau_q$: $\tau_q^{ee} \to \tau_q^{ee} + \tau_q^{dis}  $.

In this way, the effects of the electron scattering on disorder of different types,  together with the strong pair electron-electron correlations
in magnetic field, may lead to substantial change of the times $\tau_2 $ and $\tau_q$, entering the fluid motion equation (\ref{motion_eq_fin}),
and   to the relation $\tau_2 \sim \tau_q$, being  unexpected for the Fermi gas and  leading to the strongly non-monotonous
 resistance~$\varrho(\beta,\nu)$  and differential resistance~$R(\beta,\nu)$ [see Figs.~S2($a$-$d$)].

{\em 3. Comparison of theory and experiment.  }
Now let us compare the experimental result of work~\cite{non-lin_hydr_1}, in which the 2D electron magnetotransport in the non-linear regime
 in high-quality GaAsquantum wells was examined, with the predictions of our theory.

First of all, we estimate the values of the non-linearity parameter $\nu$ for realistic structures and conditions of
 experiment~\cite{non-lin_hydr_1}.

 For the high-quality GaAs quantum well samples studied in Ref.~\cite{non-lin_hydr_1} (and in
Ref.~\cite{exps_neg_4}) one can make the following estimates of characteristics of the sample and flow parameters in the maximally
strongly  non-linear regime (the current $I_{max}=350 \; \mu$A and temperature $T=1.5$~K): $ m v_F^2 \approx 2 \varepsilon _F =21 $~meV;
$ a_B = 10$~nm; $ l_2 \approx v_F \tau_2 =3 $~$\mu$m (this value was extracted in Ref.~\cite{je_visc} from the width of
the experimental magnetoresistance curves of the same sample in the linear regime, obtained in experiment~\cite{exps_neg_4},  within
the hydrodynamic framework for explanation of the giant negative magnetoresistance);  $E~0.3$~V$/$cm. The last value of the  electric field
has been estimated as $E\sim \varrho \, I_{max}/W$ from the values  of the maximal current $I_{max}= 350 $~$350\;\mu$A, the total sample
width $W=200\;\mu$m, and the mean experimental sample   resistivity~$\varrho \sim 15$~$\Omega$~\cite{exps_neg_4,non-lin_hydr_1} at magnetic
fields~$B \lesssim 1$~kG. The effective width   of the conductive channels, $W_{eff} <W$,  of  the hydrodynamic Poiseuille-like flows
was introduces   in Ref.~\cite{je_visc} in order   to fit experimental data within the hydrodynamic model. This value can be related with
 the macroscopic ``oval'' defects inside the sample bulk~\cite{d1,d1_new}, which partly block the flow.  Exactly this value $W_{eff}$, as
a characteristic of the individual conductive channel with the Poiseuille-like flow,  should in implied in Eq.~(\ref{f_,_nu}) for
 the dimensionless characteristic of the degree of non-linearity  $\nu$. For the discussed sample, it was deduced in Ref.~\cite{je_visc}
 from fitting of the experimental curves of Ref.~\cite{exps_neg_4}  that  $ W_{eff} = 10 $~$\mu$m for the discussed sample.
The result of calculation of  $\nu$ by Eq.~(\ref{f_,_nu}) with  the above  parameters is $\nu \sim 15$.  This value can correspond
to  relatively large values of the actual nonlinearity parameter  $\nu ' (\beta)= C_\nu \,f(\beta) \nu$, entering in the motion
equation~(\ref{motion_eq_fin}), those can be near the critical values of $\nu'(\beta)$ [see Fig.~S1(a,b)]. We remind that
the last values have been  obtained numerically in this work and corresponding to the situation  when the flow becomes strongly nonlinear
 and irregular, and the constructed stable laminar solution (\ref{solution}) fails.

In experiment~\cite{non-lin_hydr_1}, the differential magnetoresistance $R=dE/dI$  of high-quality GaAs quantum well samples was measured
as a function of magnetic field $B$ at different values of the current $I$ through the structure, which was the parameter controlling
the non-linearity magnitude. The observed curves $R(B;I)$ became non-monotonic with the increase of the  current in the region of
relatively low magnetic fields [see Fig.~1 in~\cite{non-lin_hydr_1}, cited in panel~($e$) of Fig.~S2]. At sufficiently high currents~$I$
the value $R(\beta)$ exhibits  specific features:  appearance of a strong minimum at $B=0$ and a maximum at some $B^*>0$. With the increase
of $I$, a smaller additional maximum at $B=0$ additionally appear. Some small features  (weak minimum and maximum, or, say, shoulders)
appear also at higher  $B$ [see Fig.~3 in~\cite{non-lin_hydr_1}, cited in panel~($f$) of Fig.~S2].

Much similarity is seen between the curves in panels~($b$) and~($e$): both experimental and theoretical curves exhibit the appearance of
 a maximum  at some $B^* >0$  with the increase of the non-linearity parameter ($\nu$ related to $I$ as it was discussed above).
Note also the on theoretical curves, at not too strongly non-linearity regime,  a ``linear-regime'' maximum at $B =0$ remains comparable
with the ``appeared non-linear-regime'' maximum at $B^* >0$. These features correlate with the appearence of a``non-linear-regime-maximum''
at $B^* >0$ on the experimental curves with the increase of~$I$.

Another point in comparison of our theory and experimental results of Ref.~\cite{non-lin_hydr_1} may be as follows.
Additional maximums and minimums at higher $B$ [panel~($f$) in Fig.~S2] in experiment may be partially related to the strong abrupt falling
of $R(\beta)$ and some critical $\beta_c$ within the discussed model.

Herewith  we note the the dependence of $ B^* $ on the non-linearity
parameter   as well as the dependencies of $R$ at $B=0$ and $B \gg B^*$ on the non-linearity parameter~$I$ are not explained within
 the considered here mechanism.

 There can be other mechanisms of the dependence of the viscosity on the velocity gradient, first of all, related to the heating mechanism:
 increase of the actual temperature of 2D electrons  by the Joule heat at sufficiently large  currents. Our preliminary estimations,
 based on calculation of  the Joule heat  power on a viscous flow and the rate of the energy relaxation of 2D electron due to the interaction
 with the acoustic phonons, evidences   that exactly the last mechanism can be responsible for the experimentally observed smooth decrease
 of the differential resistance at zero $B$ and the increase  at $B \gg B^* $ [see Fig.~S2($e$)]. We emphasize that the last effect is
 absent within the developed here memory-effect mechanism of non-linearity [see Figs.~S2($b$,$d$)].  All this issues requires further research.

In this way, a comparison of experimental differential magnetoresistance $R(B,I)$ and the results of the calculations of $R(\beta,\nu)$
within our model shows that, at least, a part of  the experimental   properties  of $R(B,I)$ can be explained by the proposed here  mechanism
 for non-linear magnetotransport due to the memory effect induced by the pair electron-electron correlations.


\begin{thebibliography}{}
\bibitem{ft}
The variables $\mathbf{E}_y ^{int} (y,t) $ and~$ \delta n (y,t)$ are determined from
 the separate equations via $ V(y,t) $.
\\
\bibitem{f2}
We emphasize that the motion equation~(\ref{motion_eq_fin}) does not account other inhomogeneous and non-linear effects,
such as the kinematic term $(\mathbf{V} \cdot \nabla )\mathbf{V}$;  the thermoelectric effects; and the heating effect, controlled
by the energy balance with phonons  and  the heat  transport (see, for reference, works~\cite{Volkov,Semina,Karpus_1,Yassievich}).
\\
\end{thebibliography}
\end{document}